\documentclass[preprint,prd,aps,floatfix,preprintnumbers,superscriptaddress,bibnotes,nofootinbib]{revtex4-1}

\usepackage{epsfig}
\usepackage{amsmath}
\usepackage{latexsym}
\usepackage[psamsfonts]{amssymb}
\usepackage{graphicx}
\usepackage{ulem}
\usepackage{longtable}
\usepackage{epstopdf}
\usepackage{bm}
\usepackage{color}

\newcommand{\be}{\begin{equation}}
\newcommand{\ee}{\end{equation}}
\newcommand{\bea}{\begin{eqnarray}}
\newcommand{\eea}{\end{eqnarray}}
\newcommand{\bi}{\begin{itemize}}
\newcommand{\ei}{\end{itemize}}

\begin{document}


\title{
On the equivalence between the Wilson flow and stout-link smearing
}


\author{Masato Nagatsuka}\email{masaton@nucl.phys.tohoku.ac.jp}
\author{Keita Sakai}\email{sakai@nucl.phys.tohoku.ac.jp}
\author{Shoichi Sasaki}\email{ssasaki@nucl.phys.tohoku.ac.jp}

\affiliation{Department of Physics, Tohoku University, Sendai 980-8578, Japan}

\date{\today}
\begin{abstract}
We present the numerical equivalence between the Wilson flow and stout-link smearing, 
both of which are known 
to be a well established
technique for smoothing the gauge fields on the lattice.
Although the conceptional correspondence between two methods 
was first pointed out by L\"uscher in his
original paper [J. High Energy Phys.~08 (2010) 071], 
we provide a direct analytical proof of the equivalence 
between the two methods at finite lattice spacing $a$ in the zero limit of the stout-smearing parameter $\rho$.
The leading order corrections start at ${\cal O}(\rho)$, which would induce ${\cal O}(a^2)$ corrections. It is, therefore, not obvious that they remain equivalent even with finite parameters 
($a\neq 0$ and $\rho\neq0$) within some numerical precision. In this paper, we demonstrate the equivalence of both methods by directly comparing the expectation
value of the action density, which is measured in actual numerical simulations.

\end{abstract}


\maketitle

 
\section{Introduction}
\label{sec:INTRO}
The Yang-Mills gradient flow has been used
extensively in lattice gauge simulations in a variety of 
aspects beyond its original proposal~\cite{{Narayanan:2006rf},{Luscher:2009eq},{Luscher:2010iy}}.
The Yang-Mills gradient flow is a kind of diffusion equation where the
gauge fields evolve smoothly as a function of fictitious time (denoted as flow time). 
One of the major benefits of the Yang-Mills gradient flow is that
the correlation functions of the flowed gauge fields for a positive flow time
become ultraviolet (UV) finite to all orders in perturbation theory without any multiplicative wave function renormalization~\cite{Luscher:2011bx}. Therefore, the gradient flow technique can serve as a renormalization scheme with help of the perturbation theory. Along this line of thought, this method is extremely useful for determining high-precision reference scale determination~\cite{{Borsanyi:2012zs},{Sommer:2014mea}}, computing the nonperturbative running of the coupling constant~\cite{Ramos:2014kla} and chiral condensate~\cite{Luscher:2013cpa}, defining the energy-momentum tensor on the lattice~\cite{Suzuki:2013gza}, and so on~\cite{Luscher:2013vga}. 

The Wilson flow that is the lattice version of 
the Yang-Mills gradient flow, makes the link variables $U_\mu(x)$
diffused in the four-dimensional space-time at finite flow time $t$. 
The associated flow $V_\mu(t,x)$ of lattice gauge fields is defined
by the following differential equation with the initial conditions
$V_\mu(0,x)=U_\mu(x)$:
%
%
\begin{equation}
\frac{\partial}{\partial t}V_\mu(t,x)=-g_0^2 \partial_{x,\mu}S_{W}[V] V_\mu(t,x),
\label{eq:Wilson_flow}
\end{equation}
where $S_{W}[V]$ denotes the standard Wilson plaquette action in terms
of the flowed link variables $V_\mu(t,x)$. The operator $\partial_{x,\mu}$ stands
for the Lie-algebra valued differential operator with respect to
the link variable~\cite{{Luscher:2010iy},{Luscher:2009eq}}. According to Eq.~(\ref{eq:Wilson_flow}), 
the diffusion process gives rise to another aspect of the associated flow that can highly suppress the local ultraviolet fluctuations on the gauge fields. 
In this sense, the gradient flow method is regarded as an alternative approach instead of the single-link smearing in both the computation of topological charge and susceptibility~\cite{{Bonati:2014tqa}, {Petreczky:2016vrs},{Alexandrou:2017hqw}} and the glueball spectroscopy~\cite{{Chowdhury:2015hta},{Sakai:2022zdc}}.

Indeed, the similarity between the Yang-Mills gradient flow and the stout-link smearing was first pointed out in the original L\"uscher's papers~\cite{{Luscher:2010iy},{Luscher:2009eq}}. 
The stout-link smearing is a well established
smearing scheme and is defined as the following recursive procedure~\cite{Morningstar:2003gk}. 
Here, for simplicity, 
the stout-smearing parameters $\rho_{\mu\nu}$ are taken as
$\rho_{\mu\nu}=\rho$. The link variables $U_{\mu}^{(k)}(x)$ at step
$k$ are mapped into the link variables $U_{\mu}^{(k+1)}(x)$ using
%
%
\begin{equation}
U_{\mu}^{(k+1)}(x)=\exp\left(
i\rho Q^{(k)}_{\mu}(x)
\right)U_{\mu}^{(k)}(x),
\label{eq:new_link_stout}
\end{equation}
where $Q^{(k)}_{\mu}(x)$ corresponds to
a Lie algebra valued quantity given by 
%
%
\begin{equation}
Q^{(k)}_{\mu}(x)=ig_0^2 \partial_{x,\mu}S_{W}[U_{\mu}^{(k)}(x)]
\label{eq:lie_algebra_value}
\end{equation}
with the Wilson action in terms of the stout link $U_{\mu}^{(k)}(x)$. 
Based on the expression of Eq.~(\ref{eq:lie_algebra_value}), which was first derived  
in Ref.~\cite{Luscher:2009eq}, L\"uscher pointed out that the Euler integration of the Wilson flow can be regarded as a continuous version of the recursive update procedure in the stout-link smearing~\cite{{Luscher:2010iy},{Luscher:2009eq}}. 

This idea was followed by Bonati and D'Elia, who carried out a systematic investigation 
on the numerical equivalence between the Wilson flow and standard cooling for the determination of topological observables~\cite{Bonati:2014tqa}. As a subsequent work, 
a comprehensive comparison was performed among the Wilson flow 
and various smearing methods including the stout-link smearing in Ref.~\cite{Alexandrou:2017hqw}. 
In these studies the classical $a$-expansion is used to lead the approximate equivalence between
the Wilson flow and various smoothing techniques, which is considered to 
appear at the relatively finer lattice spacing~\cite{{Bonati:2014tqa},{Alexandrou:2017hqw}}.

In the previous works~\cite{{Bonati:2014tqa},{Alexandrou:2017hqw}}, 
the numerical equivalence between the Wilson flow
and smoothing techniques are demonstrated for calculation of the topological observables. 
The topological observables should be insensitive to the true quantum fluctuations, though the measurement of the topological charge on the lattice is highly sensitive 
to the UV fluctuations caused by the lattice artifacts. There is no theoretical warrant 
for keeping the true quantum fluctuations out of the UV fluctuations. Therefore, the situation can be non-trivial for other observables, which are associated with the long distance physics and also fully responsible for the quantum fluctuation survived in the continuum. The most simple example for the quantum observable is the expectation value of the action density~\footnote{
The numerical correspondence between the Wilson flow and stout-link smearing has been reported  without detailed discussion 
in Ref.~\cite{Thomas:2014tda} similar to the one observed between the Wilson flow and standard cooling in Ref.~\cite{Bonati:2014tqa}.},  which is indeed a key ingredient in the gradient flow 
approach to determine the high-precision reference scale. 

Recently, two of our collaborators had found that there is numerical equivalence between the spatial Wilson flow~\footnote{The diffusion is restricted only to spatial directions.} and the stout-link smearing in the glueball spectroscopy~\cite{{Sakai:2022zdc},{Sakai:2022mgd}}. 
To understand their numerical results, an analytical proof of equivalence between the Wilson flow and the stout-link smearing was also revisited in Ref.~\cite{Sakai:2022zdc}.
Although, two of our collaborations gave a derivation of the Wilson-flow differential equation from the stout-link smearing in Ref.~\cite{Sakai:2022zdc}, there is some apparent
error~\footnote{
In Eq.~(A.10) of Ref.~\cite{Sakai:2022zdc}, the higher-order terms involving iterated commutators of 
$iQ_\mu^{(k)}(x)$ and $\log{U^{(k)}_\mu(x)}$ that do not vanish even in the limit of $\rho\to 0$ are omitted.
We thank M. Ammer for pointing out this apparent error~\cite{Ammer} in our proof represented in Ref.~\cite{Sakai:2022zdc}.}
leading to the fact that the equivalence of the two methods is satisfied only at the zero limits of both 
the lattice spacing $a$ and the stout-smearing parameter $\rho$. 
One correcting the error in the previous proof presented in Ref.~\cite{Sakai:2022zdc}, we can explicitly derive a continuous version of the stout smearing procedure {\it at finite lattice spacing $a$} in the limit of $\rho\to0$ as will be discussed in Sec.~\ref{sec:FLOW_STOUT}. 
However, it is still not obvious that the two methods remain equivalent even with finite parameters within some numerical precision. 
Therefore, in this paper, we would like to verify the equivalence of
the two methods by measuring the expectation value of the action density $\langle E\rangle$ (see Sec.~\ref{sec:referece_scales} for definition) used to determine the Wilson flow scales $t_0$ and $\omega_0$, in numerical simulations. 

This paper is organized as follows. 
In Sec.~\ref{sec:FLOW_STOUT}, we give a short outline for the
analytical proof of the equivalence between 
the Wilson flow and the stout smearing. 
Section~\ref{sec:Num_results} presents the numerical results
of the action density given by both the Wilson flow and the stout-smearing methods, 
and then directly compare two results
to verify the satisfactory condition of the
equivalence with respect to the finite lattice spacing $a$ and the finite stout-smearing parameter $\rho$. 
Finally, we close with summary in Sec.~\ref{sec:SUMMARY}.

\clearpage
\section{Equivalence between the Wilson flow and the stout-link smearing}
\label{sec:FLOW_STOUT}

As described in Sec.~\ref{sec:INTRO}, the Wilson flow equation~(\ref{eq:Wilson_flow}) and
the stout-link smearing procedure~(\ref{eq:new_link_stout}) are connected through
the relation~(\ref{eq:lie_algebra_value}) that was first derived in Ref.~\cite{Luscher:2009eq}.
Therefore, the gradient flow can be regarded as a continuous version of the recursive 
update procedure in the stout-link smearing as pointed out in the original papers of the Wilson flow~\cite{{Luscher:2010iy},{Luscher:2009eq}}.
In this section, we elaborate on an analytical derivation of the Wilson-flow differential equation from the stout-link smearing in line with Ref.~\cite{Sakai:2022zdc}
as an extension of Refs.~\cite{{Luscher:2009eq},{Bonati:2014tqa}}.

For this purpose, let us first introduce a continuous variable $s$ associated with smearing step $k$. Therefore, the stout link $U_{\mu}^{(k)}(x)$ and the associated operator $Q_{\mu}^{(k)}(x)$
at step $k$ will be represented by functions of the continuous variable $s$ denoted as $\tilde{U}_\mu(s,x)$ and $Q_\mu(s,x)$.

\subsection{Derivation of the flow equation from the stout smearing}

Our derivation process  
consists of the following two steps:
\begin{enumerate}
\item Derive a continuous version of the stout-smearing procedure given below:
%
%
\begin{equation}
\frac{\partial}{\partial s}\tilde{U}_\mu(s,x)=iQ_\mu(s,x) \tilde{U}_\mu(s,x).
\label{eq:stout_flow_equation}
\end{equation}

\item Derive the explicit form of the link derivative of the Wilson gauge action $S_W$ given below~\cite{{Luscher:2009eq},{Bonati:2014tqa}}:
%
%
\begin{equation}
g_0^2\partial_{x,\mu}S_W[\tilde{U}]=-iQ_\mu(s,x).
\label{eq:DeriveAction_cont}
\end{equation}

\end{enumerate}
which are presented in the proceeding two subsections:
Sec.~\ref{sec:step1_proof} and Sec.~\ref{sec:step2_proof}.

\subsubsection{Step 1: the differential equation for the stout smearing}
\label{sec:step1_proof}

Let us derive a continuous version of the stout-smearing procedure.
By subtracting $U^{(k)}_\mu(x)$
from both sides of Eq.~\eqref{eq:new_link_stout},
%
%
\begin{equation}
\Delta_k U^{(k)}_\mu(x)=\left(\exp\left(i\rho Q_{\mu}^{(k)}(x)\right)-1\right)U^{(k)}_\mu(x), 
\label{eq:derivative_smear}
\end{equation}
where ${\Delta_k}$ represent a forward difference with respect to $k$ is defined by 
%
%
\begin{equation}
{\Delta_k}F(k)\equiv F(k+1)-F(k).
\end{equation}
If a continuous variable ${s=\rho k}$ is introduced, by performing the variable conversion
on $F$ as $F(k)=\tilde{F}(s)$, the difference of $F$ becomes the derivative of $\tilde{F}$ with respect to
$s$ in the limit of $\rho \to 0$ as below:
%
%
\begin{equation}
  \lim_{\rho\to 0}\frac{\Delta_k F(k)}{\rho} 
  = \lim_{\rho\to 0}\frac{\tilde{F}(s+\rho)-\tilde{F}(s)}{\rho} =\frac{\partial}{\partial s}\tilde{F}(s).
\end{equation}
To consider the limit of $\rho\to 0$ for the stout smearing parameter, 
Eq.~\eqref{eq:derivative_smear} can be rewritten as
%
%
\begin{align}
\lim_{\rho\to0}\frac{\Delta_k U^{(k)}_\mu(x)}{\rho}=&\lim_{\rho\to0}\frac{1}{\rho}\left(\exp\left(i\rho Q_{\mu}^{(k)}(x)\right)-1\right)U^{(k)}_\mu(x) \nonumber \\
=&\lim_{\rho\to0}
iQ_\mu^{(k)}
\left(
\sum_{n=0}^{\infty}\frac{(i\rho Q_\mu^{(k)})^n}{(n+1)!}
\right)U^{(k)}_\mu(x),
\end{align}
which leads to the following differential equation
%
%
\begin{equation}
  \label{eq:derivative_Uk}
  \frac{\partial}{\partial s}\tilde{U}_\mu(s, x)=iQ_\mu(s,x)\tilde{U}_\mu(s, x),
\end{equation}
where the replacements of ${U^{(k)}_\mu(x)=\tilde{U}_\mu(s, x)}$ and ${Q^{(k)}_\mu(x)=Q_\mu(s, x)}$ are used for the variable conversion ${s=\rho k}$. 
Recall that for the finite smearing parameter $\rho$,
the leading order corrections on Eq.\eqref{eq:derivative_Uk} start at ${\cal O}(\rho)$.

\subsubsection{Step 2: the link derivative of the Wilson action}
\label{sec:step2_proof}

Let us consider the explicit form of the link derivative of the Wilson gauge action $\partial_{x,\mu}S_{W}$ which appears in the Wilson flow equation~\eqref{eq:Wilson_flow}. 
Although the final expression as given in Eq.~(\ref{eq:DeriveAction_cont}) was originally derived by L\"uscher in Ref.~\cite{Luscher:2009eq}, its derivation was elaborated in Ref.~\cite{Bonati:2014tqa}.

At first, we introduce the antihermitian traceless $N\times N$ matrices $T^a$ $(a=1, ..., N^2-1)$ as generators of $SU(N)$ group~\footnote{In this paper, we use the notational conventions adopted in the original L\"uscher's paper~\cite{Luscher:2010iy}. Namely, they are normalized by ${\rm Tr}\left(T^aT^b\right)=-\frac{1}{2}\delta^{ab}$ and also satisfy the commutation relations $[T^a, T^b]=f_{abc}T^c$ with the structure constants $f^{abc}$.}.
In general, with respect to a basis $T^a$, the elements $M$ of the Lie algebra of $SU(N)$ are given 
by $M=M^aT^a$ with real components $M^a$. 
Therefore, the link derivative operator $\partial_{x,\mu}$ can be expressed with respect to a basis $T^a$ as

%
%
\begin{equation}
\partial_{\mu, x}=T^a\partial^a_{\mu, x},
\end{equation}
where the operators $\partial^a_{\mu, x}$ are defined by
%
%
\begin{equation}
\partial^a_{\mu, x}f(U)=\frac{d}{ds}f(e^{sX^a}U)|_{s=0}
\end{equation}
with
%
%
\begin{equation}
X^a(y,\nu)=
\begin{cases}
T^a & {\rm if}\ (y,\nu)=(x,\mu), \\
0 & {\rm otherwise}
\end{cases}
\end{equation}
and act as differential operators on functions $f$ of the link variable $U$.

When the link derivative $\partial_{x,\mu}$ acts on the action, 
we may simply focus on the term 
that explicitly depends on $U_\mu(x)$ in the action as
%
%
\begin{equation}
  S_{W}[U] = -\frac{2}{g_0^2}\sum_{x, \mu>\nu}\left[
   {\rm Re}{\rm Tr}\left\{U_\mu(x)\Sigma^\dagger_{\mu}(x)
   \right\}+\left\{\mbox{terms independent of $U_\mu(x)$}\right\}\right],
    \label{eq:WilsonAction2}
\end{equation}
%
where the sum of all staples neighboring $U_{\mu}(x)$ is represented 
by $\Sigma_{\mu}(x)$, which is given by
%
%
\begin{equation}
\Sigma_{\mu}(x)=\sum_{\mu>\nu}\left[U_{\nu}(x)U_{\mu}(x+\hat{\nu})U_{\nu}^\dagger(x+\hat{\mu})
+U_{\nu}^\dagger(x-\hat{\nu})U_{\mu}(x-\hat{\nu})U_{\nu}(x-\hat{\nu}+\hat{\mu})
\right].
\end{equation}
If we set $\Omega_{\mu}(x)=\Sigma_\mu(x)U^\dagger_\mu(x)$, each basis component is given as
%
%
\begin{equation}
    g_0^2\partial^a_{\mu, x}S_{W}[U]=-2{\rm Re}{\rm Tr}\left\{T^a \Omega^\dagger_\mu(x)\right\}=-{\rm Tr}\left\{T^a\left(\Omega^\dagger_\mu(x)-\Omega_\mu(x)\right)\right\},
\end{equation}
%
where $\Omega^\dagger_\mu(x)$ denotes the sum of all plaquettes that include $U_{\mu}(x)$. 
Therefore, we finally get
%
%
\begin{equation}
g_0^2\partial_{\mu, x}S_{W}[U]=-iQ_{\mu}(x)
\label{eq:DeriveAction}
\end{equation}
with 
%
%
\begin{equation}
Q_{\mu}(x)=\frac{i}{2}\left(\Omega^\dagger_\mu(x)-\Omega_\mu(x)\right)
-\frac{i}{2N}{\rm Tr}\left(\Omega^\dagger_\mu(x)-\Omega_\mu(x)\right), 
\label{eq:StoutPart}
\end{equation}
which becomes a Lie algebra valued quantity~\cite{{Luscher:2009eq},{Bonati:2014tqa}}. 
If the link variables $U_{\mu}(x)$ are
replaced by the stout links $U_{\mu}^{(k)}(x)$ (or $\tilde{U}_{\mu}(s,x)$), 
the corresponding operators of $Q^{(k)}_\mu(x)$ (or $Q_{\mu}(s,x)$) can be defined in terms of $U_{\mu}^{(k)}(x)$ (or $\tilde{U}_{\mu}(s,x)$). Thus, one can get Eq.~\eqref{eq:lie_algebra_value} (or Eq.~\eqref{eq:DeriveAction_cont}).

\subsection{Equivalence without classical $a$-expansion}

By combining with Eq.~\eqref{eq:stout_flow_equation} and Eq.~\eqref{eq:DeriveAction_cont}, we get
%
%
\begin{equation}
\label{eq:stout_differential_eq}
\frac{\partial}{\partial s}\tilde{U}_\mu(s,x)=-g^2_0\partial_{x,\mu}S_{W}[\tilde{U}]\tilde{U}_\mu(s,x),
\end{equation}
which coincides with the Wilson flow equation~\eqref{eq:Wilson_flow} 
under the correspondences of $t \leftrightarrow s$ and $V_{\mu}(t,x) \leftrightarrow \tilde{U}_{\mu}(s,x)$.
Since the variable $s=k\rho$ directly corresponds to the Wilson flow time $t$, the perturbative matching relation of the flow time $t$, and the smearing parameters $\rho$ and $n_{\rm st}$ as $t=\rho n_{\rm st}$ found in Refs.~\cite{{Thomas:2014tda},{Alexandrou:2017hqw}} is also rigorously proved {\it without classical $a$-expansion}.

We confirm that the differential equation derived from the stout smearing procedure in the limit of $\rho\to 0$ as shown in Eq.~\eqref{eq:stout_differential_eq} exactly reduces to the Wilson flow equation~\eqref{eq:Wilson_flow} 
even at finite lattice spacing $a$. When $S_W$ is replaced by the spatial part of the standard Wilson action and $\rho_{\mu}$ is set to be isotropic 3-dimensional one ($\rho_i=\rho$ and $\rho_4=0$), the equivalence between the spatial Wilson flow and the spatial stout smearing remains unchanged as was justified for the glueball spectroscopy in Ref.~\cite{{Sakai:2022mgd},{Sakai:2022zdc}}. 
For the finite smearing parameter $\rho$, the leading order corrections on Eq.~\eqref{eq:stout_differential_eq} 
start at ${\cal O}(\rho)$, which induces ${\cal O}(a^2)$ corrections since the flow time $t=\rho n_{\rm st}$ has dimension length squared~\cite{Luscher:2010iy}.

It should be emphasizing that a continuous version of the stout-smearing procedure 
can be derived as the differential equation~(\ref{eq:stout_differential_eq})
{\it without classical $a$-expansion}. Thus, the equivalence between the Wilson flow and stout-link smearing is led to be valid even at finite lattice spacing as long as the smearing parameter $\rho$ is
taken to be sufficiently small. This brings a new perspective on the application of equivalence
as discussed in the next section. 

\section{Numerical results} 
\label{sec:Num_results}

\subsection{Lattice setup}
We perform the pure Yang-Mills lattice simulations using the
standard Wilson plaquette action with a fixed physical volume
($La\approx 2.4$ fm) at four different gauge couplings ($\beta=6/g_0^2=
5.76$, 5.96, 6.17, and 6.42). 
Three of four ensembles ($\beta=5.96$, 6.17, and 6.42) (which correspond 
to the same lattice setups as in the original work of the Wilson 
flow done by L\"uscher~\cite{Luscher:2010iy}) had been generated 
for our previous study of tree-level improved lattice gradient flow~\cite{Kamata:2016any}.
In this study, we additionally generate a coarse lattice ensemble at $\beta=5.76$. 
The gauge configurations in each simulation are separated by
$n_{\rm update}$ sweeps after $n_{\rm therm}$ sweeps for thermalization
as summarized in Table~\ref{tab:Sims}.
Each sweeps consists of one heat bath~\cite{Cabibbo:1982zn} combined with four over-relaxation~\cite{Creutz:1987xi} steps. The number of configurations analyzed is 100
in each ensemble. All lattice spacings are set by the Sommer scale ($r_0=0.5$ fm). 

%
%
\begin{table*}[ht]                                                    
  \caption{
  Summary of the gauge ensembles: gauge coupling, lattice size ($L^3\times T$), plaquette value, 
  lattice spacing ($a$), spatial extent ($La$), the Sommer scale ($r_0$), 
  the number of the gauge field configurations ($N_{\rm conf}$),  
  the number of thermalization sweeps ($n_{\rm therm}$)  
  and the number of update sweeps ($n_{\rm update}$).
  All lattice spacings are set by the Sommer scale ($r_0=0.5$ fm)~\cite{{Sommer:1993ce},{Necco:2001xg}}.
  \label{tab:Sims}
  }
\begin{ruledtabular}                                                                   
\begin{tabular}{ c c c c c c c c c }
\hline
$\beta=6/g_0^2$ & $L^3\times T$ & plaquette & $a$ [fm] & $La$ [fm] & $r_0/a$ (Ref.~\cite{Necco:2001xg}) & $N_{\rm conf}$ & $n_{\rm therm}$ & $n_{\rm update}$ \cr 
\hline
5.76 & $16^3\times 32$ & 0.560938(9)     & 0.1486(7)    & 2.38   & 3.364(17)     & 100 & 5000 & 200 \cr
5.96 & $24^3\times 48$ & 0.589159(3)    & $0.1000(5)$  & 2.40   & 5.002(25) & 100 & 2000 & 200 \cr
6.17 & $32^3\times 64$ & 0.610867(1)    & $0.0708(3)$  & 2.27   & 7.061(35) & 100 & 2000 & 200\cr
6.42 & $48^3\times 96$ & 0.632217(1)    & $0.0500(2)$  & 2.40   & 10.00(5)  & 100  & 2000 & 200\cr
\hline
\end{tabular}
\end{ruledtabular}                                                                                
\end{table*}

\subsection{Wilson flow reference scales}
\label{sec:referece_scales}

We will later determine two types of the Wilson flow scales $t_0$~\cite{Luscher:2010iy} and $\omega_0$~\cite{Borsanyi:2012zs} using the expectation value of the clover-type action density $E(t, x) =\frac{1}{2}
{\rm Tr}\{G^{\rm cl}_{\mu\nu}(t, x)G^{\rm cl}_{\mu\nu}(t, x)\}$. 
The clover-leaf operator $G^{\rm cl}_{\mu\nu}(t, x)$~\cite{Luscher:2010iy} is defined by
%
%
\begin{eqnarray}
G^{\rm cl}_{\mu\nu}(t, x)&=&\frac{1}{4}\left(
V_\mu(t,x)V_\nu(t,x+\hat{\mu})V^{\dagger}_\mu(t,x+\hat{\nu})V^{\dagger}_\nu(t,x)\right.\cr
&& + V_{\nu}^{\dagger}(t,x-\hat{\nu})V_\mu(t,x-\hat{\nu})V_\nu(t,x+\hat{\mu}-\hat{\nu})V^{\dagger}_\mu(t,x)\cr
&& + V^{\dagger}_{\mu}(t,x-\hat{\mu})V^{\dagger}_{\nu}(t,x-\hat{\mu}-\hat{\nu})
V_{\mu}(t,x-\hat{\mu}-\hat{\nu})V_{\nu}(t,x-\hat{\nu})\cr
&&\left.
+V_{\nu}(t,x)V^{\dagger}_{\mu}(t,x-\hat{\mu}+\hat{\nu})V^{\dagger}_{\nu}(x-\hat{\mu})
V_{\mu}(t,x-\hat{\mu})
\right)_{\rm AH},
\end{eqnarray}
where we have introduced the traceless antihermitian projection defined as
%
%
\begin{equation}
(M)_{\rm AH}=\frac{1}{2}(M-M^\dagger)-\frac{1}{2N}{\rm Tr}(M-M^\dagger)
\end{equation}
for an $N\times N$ matrix $M$ in $SU(N)$ color space.

The vacuum expectation of $E(t, x)$ in small-$t$ regime, where the gauge
coupling becomes small, can be evaluated in perturbation theory. 
In the original paper of L\"uscher~\cite{Luscher:2010iy},
$\langle E\rangle$ was given at the next-to-leading order (NLO) in
powers of the renormalized coupling in the $\overline{\rm MS}$ scheme, while
its next-to-NLO (NNLO) correction has been evaluated by Harlander and Neumann~\cite{Harlander:2016vzb}. 

The dimensionless combination $t^2\langle E(t)\rangle$ is expressed in terms of the $\overline{\rm MS}$ running coupling $g$ at a scale of $q=1/\sqrt{8t}$ for the pure SU(3) Yang-Mills theory:
%
%
\begin{equation}
\label{eq:flowed_energy}
t^2 \langle E (t) \rangle = \frac{3 g^2(q)}{16 \pi ^2} \Bigl[ 1 
+ \frac{k_1}{4\pi} g^2(q) +\frac{k_2}{(4\pi)^2} g^4(q)+  {\mathcal O}(g^6(q)) \Bigr],
\end{equation}
where the NLO coefficient $k_1$ was obtained analytically as $k_1=1.0978$~\cite{Luscher:2010iy}, 
while the NNLO coefficient $k_{2}$ has been evaluated with the aid of numerical integration as $k_2=-0.982$~\cite{Harlander:2016vzb}. 

The lattice version of $t^2 \langle E (t) \rangle$ obtained 
in numerical simulations shows a monotonically increasing behavior as a function of the flow time $t$ and also good scaling behavior with consistent values of the continuum perturbative calculation 
(\ref{eq:flowed_energy}) that suggests the presence of the proper continuum limit~\cite{Luscher:2010iy}. 

The observed properties of $\langle E (t)\rangle$
offer a new reference scale $t_0$, is given by the solution of the following equation~\cite{Luscher:2010iy}
%
%
\begin{equation}
\left. t^2 \langle E(t) \rangle \right|_{t=t_0}= 0.3.
\label{eq:new_scale}
\end{equation}
In addition, an alternative reference scale 
was also proposed by the BMW collaboration~\cite{Borsanyi:2012zs}, as below
%
%
\begin{equation}
\left.t\frac{d}{dt}t^2\langle E(t)\rangle\right|_{t=w_0^2}=0.3,
\end{equation}
where the derivative operation of $t\frac{d}{dt}$ can 
suppress the lattice discretization effect on $t^2\langle E(t)\rangle$ in the small-$t$ regime.

In this study, the forth-order Runge-Kutta scheme is used for the Wilson flow with three integration step
sizes of $\epsilon=0.1$, 0.025, and 0.01. The flow time $t$ is
given by $\epsilon \times n_{\rm flow}$ where $n_{\rm flow}$
denotes the number of flow iterations.

As shown in Table~\ref{tab:CompWilsonFlow}, the results of $t_{0}/a^{2}$ are reasonably converged when the integration step size is smaller than 0.025. We have also checked our code of the Wilson flow by determining a reference scale of $t_{0}/a^{2}$ with
the clover-type action density, which can be directly compared with the results of Ref.~\cite{Luscher:2010iy}, as tabulated in Table~\ref{tab:CompWilsonFlow}. 
Hereafter, we use the Wilson flow results for $\epsilon=0.025$ to compare with the stout link smearing results.

%
%
\begin{table*}[ht]
\caption{
Results of $t_{0}/a^2$ obtained from the Wilson flow with three integration steps $\epsilon=0.01,~0.025,~0.001$
in comparison with the original work of L\"uscher~\cite{Luscher:2010iy}.
\label{tab:CompWilsonFlow}
}
\begin{ruledtabular} 
\begin{tabular}{lccccccc}
\hline 
&&&\multicolumn{3}{c}{$t_{0}/a^2$ (ours)} & $t_{0}/a^2$ (L\"uscher) \cr
 \cline{4-6}
 $\beta$ & $L^3 \times T$ & Statistics 
 &$\epsilon=0.1$ & $\epsilon=0.025$ & $\epsilon=0.01$ & $\epsilon=0.01$ \cr
\hline
5.76 & $16^3\times 32$&  100 & 1.2730(31) & 1.2741(31) & 1.2742(31) & N/A\cr
5.96 & $24^3\times 48$&  100 & 2.7962(62) & 2.7968(62) & 2.7968(62) & 2.7854(62)\cr
6.17 & $32^3\times 64$&  100 & 5.499(13)  & 5.499(13)  & 5.499(13)  & 5.489(14)\cr
6.42 & $48^3\times 96$&  100 & 11.242(23) & 11.242(23) & 11.242(23) & 11.241(23)\cr
\hline
\end{tabular}
\end{ruledtabular}
\end{table*}

\subsection{Comparison of the stout-link smearing to the Wilson flow}

%
%
\begin{figure*}[ht]
\centering
\includegraphics*[width=.48\textwidth]{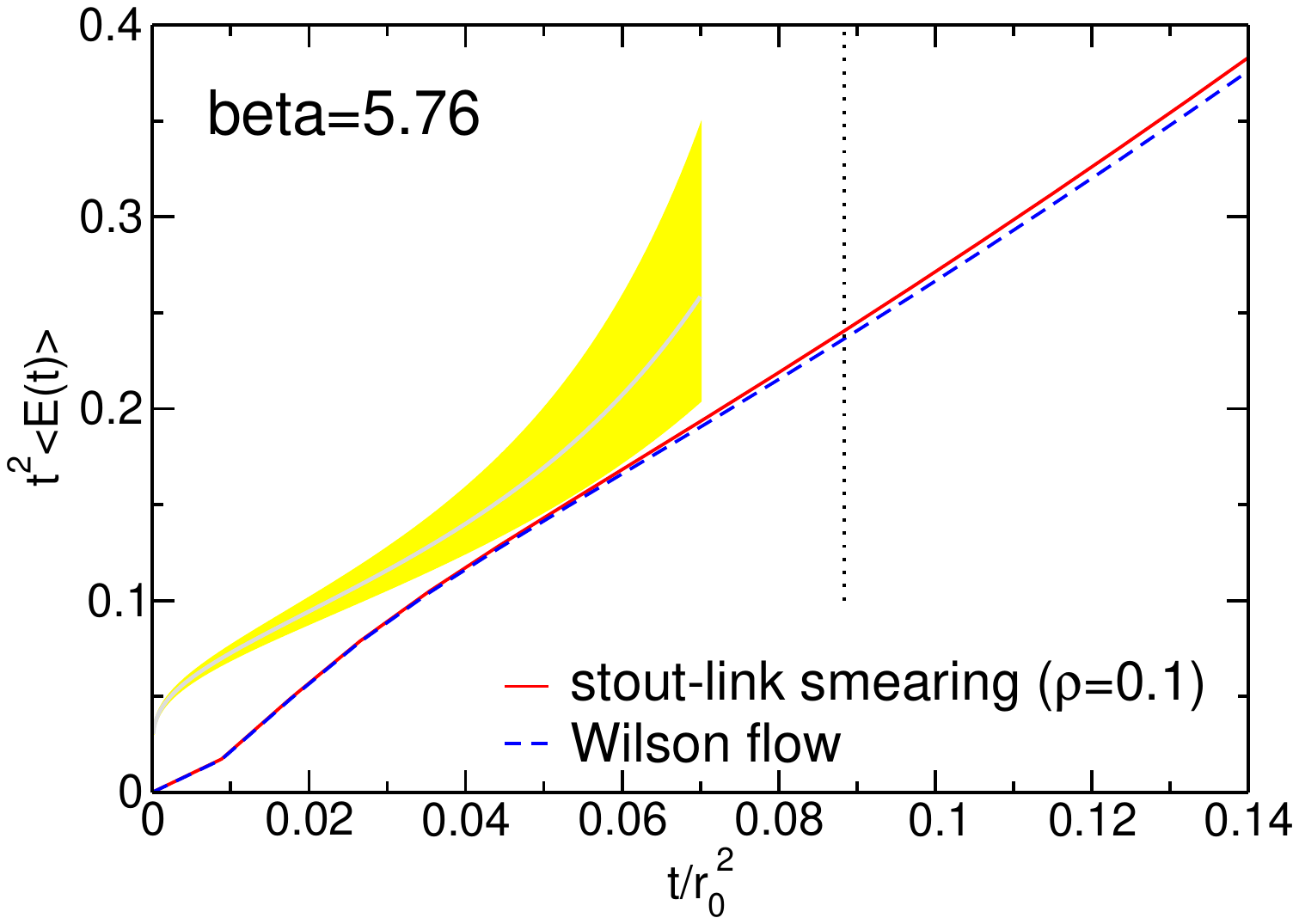} 
\includegraphics*[width=.48\textwidth]{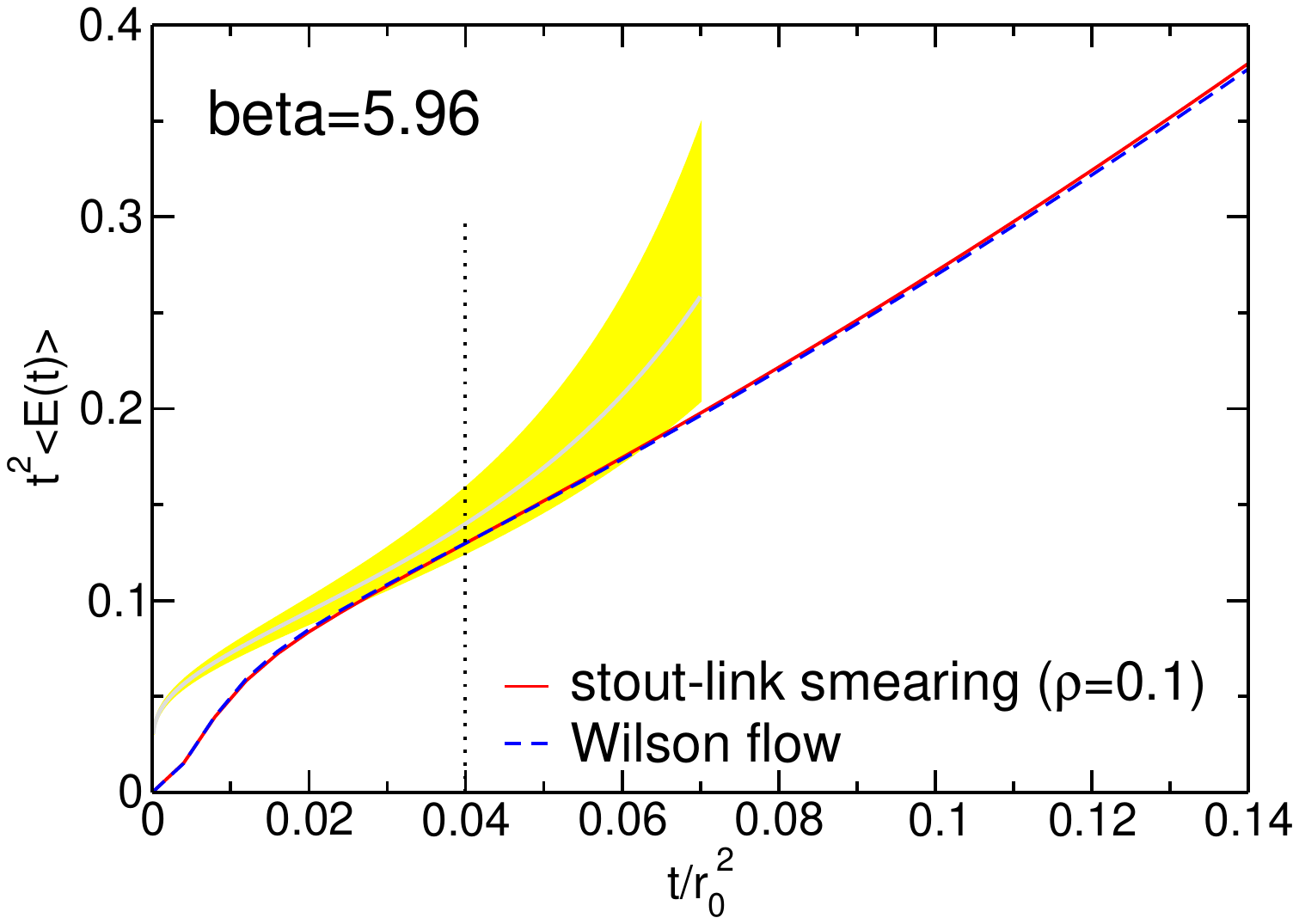} 
\includegraphics*[width=.48\textwidth]{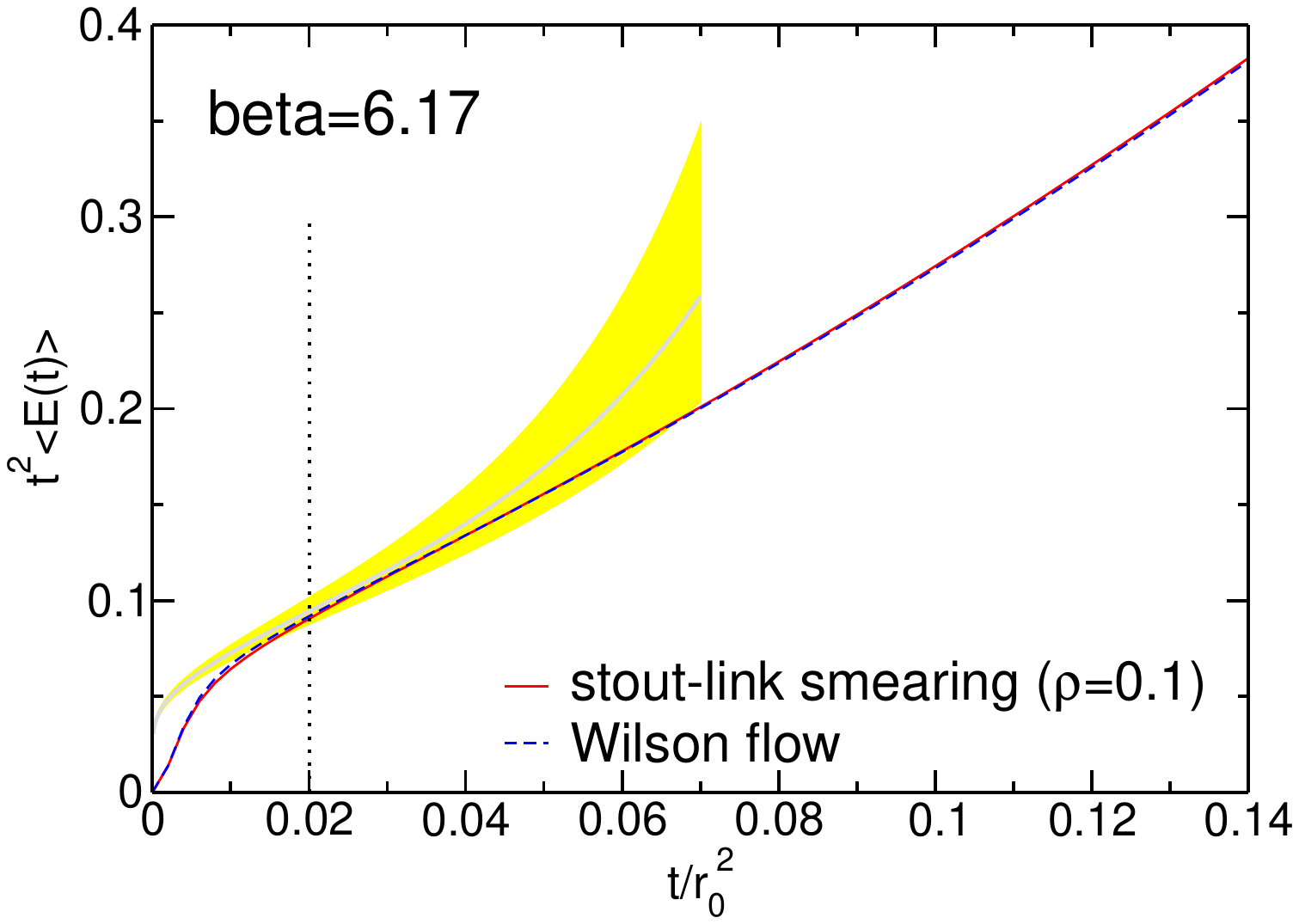} 
\includegraphics*[width=.48\textwidth]{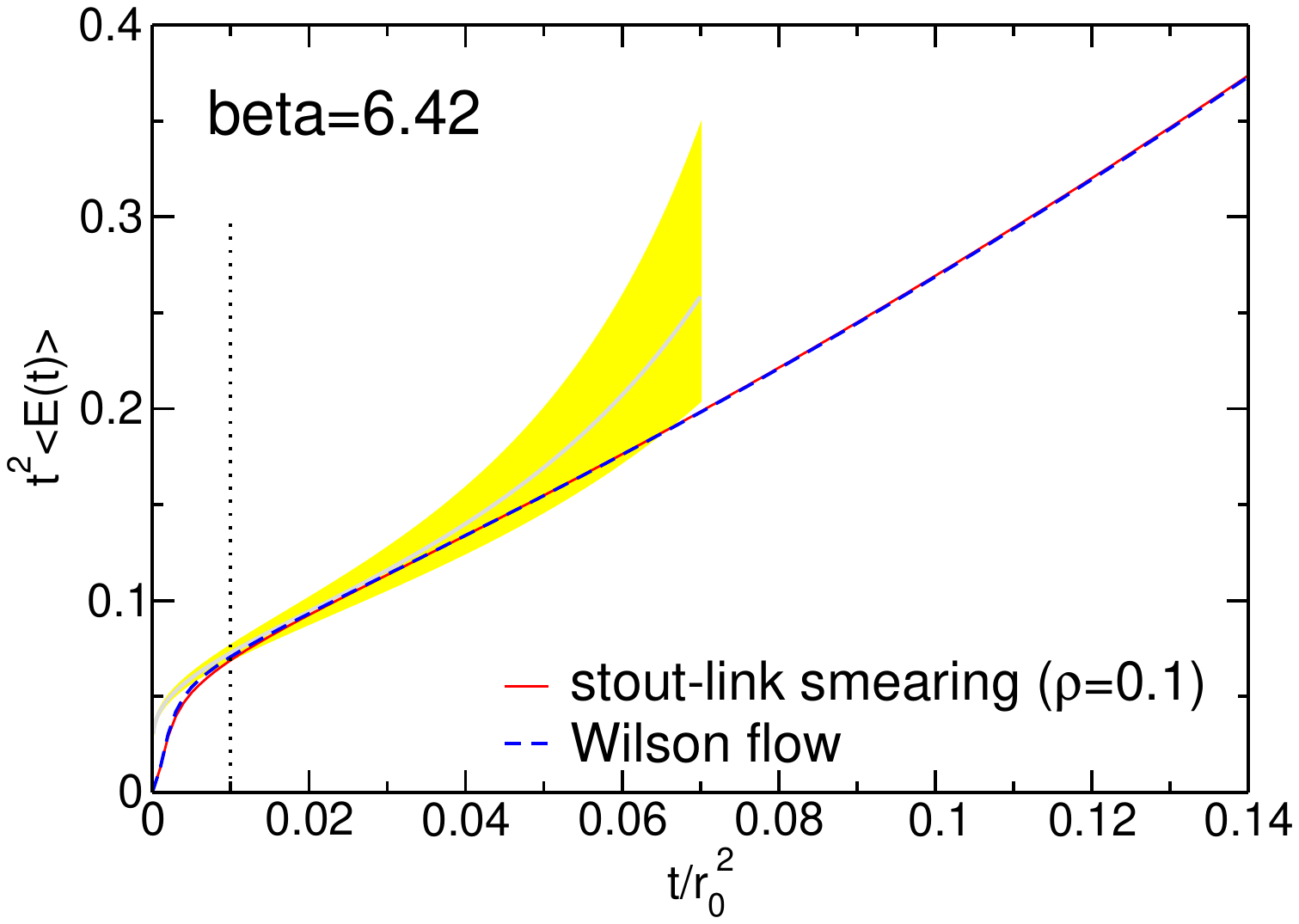} 
  \caption{The behavior of $t^2\langle E(t)\rangle$ computed by the Wilson flow (blue dotted curve) and the stout-link smearing with $\rho=0.1$ (red solid curve) at $\beta=5.76$ (upper-left panel),  $\beta=5.96$ (upper-right panel), $\beta=6.17$ (lower-left panel), and $\beta=6.42$ (lower-right panel).
  The gray solid curve with the yellow band corresponds to the continuum perturbative calculation~\cite{Luscher:2010iy} in each panel.}
\label{fig:WG}
\end{figure*}

We perform the stout smearing with $\rho=0.1$, 0.025, and 0.01 for each ensemble
listed in Table~\ref{tab:Sims} to evaluate $X(t)\equiv t^2 \langle E(t) \rangle$, where $t$ indicates the flow time given by
the matching relation of $t=\rho n_{\rm st}$.
We also evaluate $X(t)$ with the Wilson flow with a fixed integration step size of $\epsilon=0.025$. Figure~\ref{fig:WG} shows typical behaviors of $X(t)$ calculated by both the Wilson flow (blue dotted curve) and the stout-link smearing with the smearing parameter $\rho=0.1$ (red solid curve) for $\beta=5.76$ (upper-left panel), $\beta=5.96$ (upper-right panel), $\beta=6.17$ (lower-left panel), and $\beta=6.42$ (lower-right panel). The vertical dotted line 
is marked at the position of $t/r_0^2=a^2/r_0^2$ in each panel. 

In the small-$t$ region, corresponding the left side of that vertical line, 
the lattice discretization errors on $t^2\langle E(t)\rangle$ are considered non-negligible, since the lattice spacing dependence of the tree-level contribution is classified by powers of $a^2/t$~\cite{Fodor:2014cpa}. Therefore, the numerical results for the case of the finer lattice spacing show better overlap with the result of the continuum perturbative calculation displayed as the gray solid curve with the yellow band in each panel. 
Surprisingly, at a glance, the stout-link smearing procedure can well reproduce
the behavior of $t^2\langle E(t)\rangle$ that was obtained by the Wilson flow
even at the coarse lattice spacing ($\beta=5.76$) where the Wilson flow result certainly
deviates from the perturbative calculation of the continuum Yang-Mills gradient flow.

If one takes a closer look, the stout-smearing result obtained at $\beta=5.76$ is slightly deviated from the curve of the Wilson flow result. This deviation tends to
become larger as the flow time increases. However, for the cases ($\beta \ge 5.96$),
the red solid and blue dotted curves almost overlap each other and then are
visually indistinguishable.

\begin{figure*}[ht]
\centering
\includegraphics*[width=.8\textwidth]{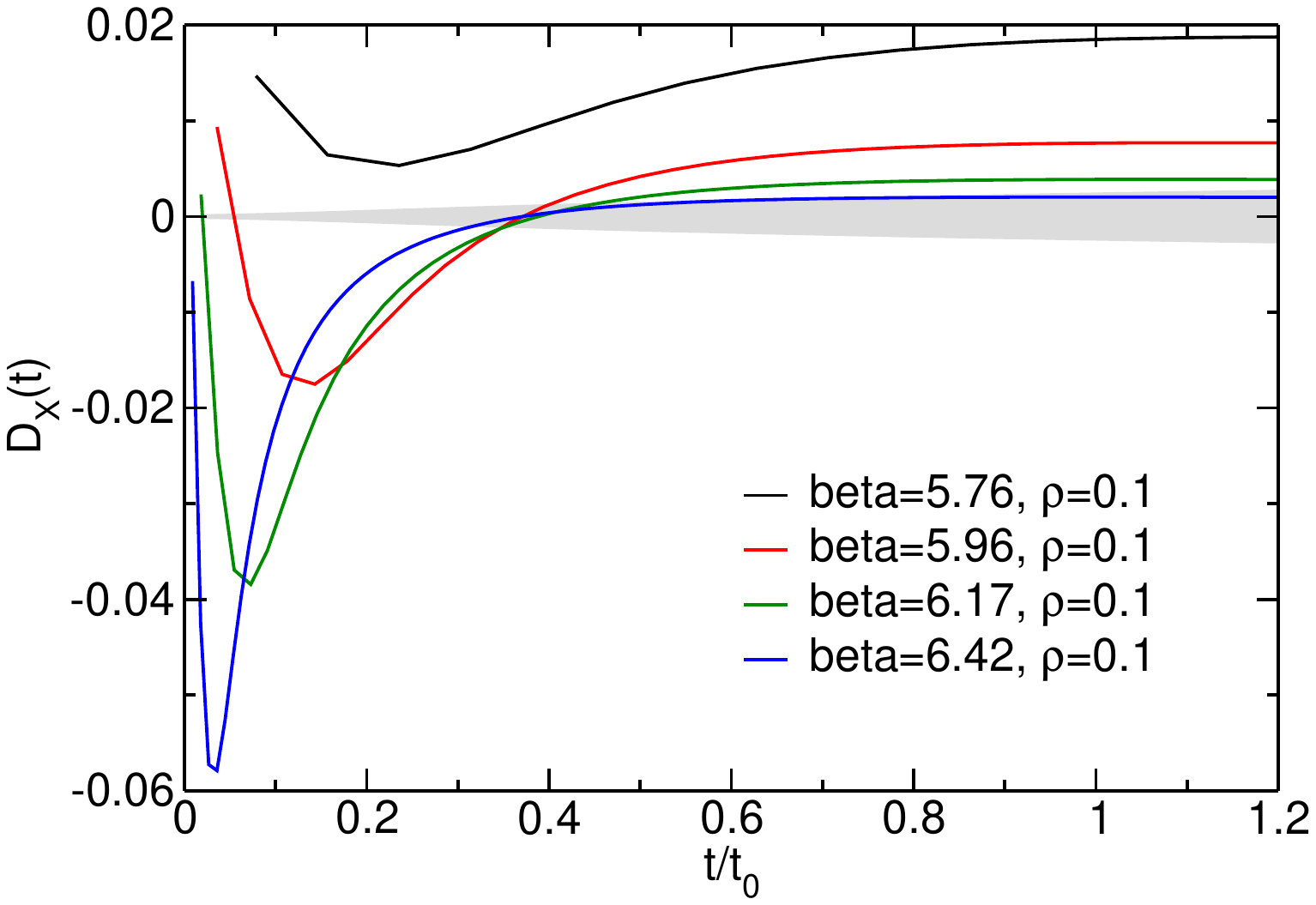} 
\caption{
The lattice spacing dependence of $D_X(t)$ calculated between the Wilson flow and the stout smearing ($\rho=0.1$) at $\beta=5.76$, 5.96, 6.17, and 6.42.
The area shaded in gray corresponds to the relative size of the statistical uncertainties
on $X(t)$ determined by the Wilson flow in this study.}
\label{fig:X_r01}
\end{figure*}

To discuss the differences in more detail, we calculate the following ratio:
%
%
\begin{equation}
D_X(t)\equiv\frac{X_{\rm stout}(t)-X_{\rm flow}(t)}{X_{\rm flow}(t)},
\end{equation}
where $X_{\rm stout}(t)$ denotes the corresponding quantity of $X(t)$ 
calculated by using the stout-link smearing as a function of
$t=\rho n_{\rm st}$, while $X_{\rm flow}(t)$
is calculated by the Wilson flow. The quantity of $D_X(t)$ can
expose the relative difference on the dimensionless quantity of $t^2\langle E(t)\rangle$ obtained from the two methods.
Figure~\ref{fig:X_r01} shows the behavior of $D_X(t)$ as a function of $t/t_0$ for all cases of four $\beta$ values with fixed smearing parameter of $\rho=0.1$. 
Although some peak structure is commonly observed in the small-$t$ region ($t/t_0< a^2/t_0$) for the finer lattice spacing ($\beta\ge 5.96$), the relative differences
are saturated in the large-$t$ region ($t/t_0 > a^2/t_0$) for all cases. 
As clearly seen in Fig.~\ref{fig:X_r01}, the saturated value of $D_X(t)$ becomes smaller as the lattice spacing decreases as expected from what we learned
in Sec.~\ref{sec:FLOW_STOUT}. It is worth stressing that the saturated difference
at $t=t_0$ becomes smaller than the statistical uncertainties on $X(t)$ given by the Wilson flow, that are represented by the area shaded in gray.

\begin{figure*}[ht]
\centering
\includegraphics*[width=.8\textwidth]{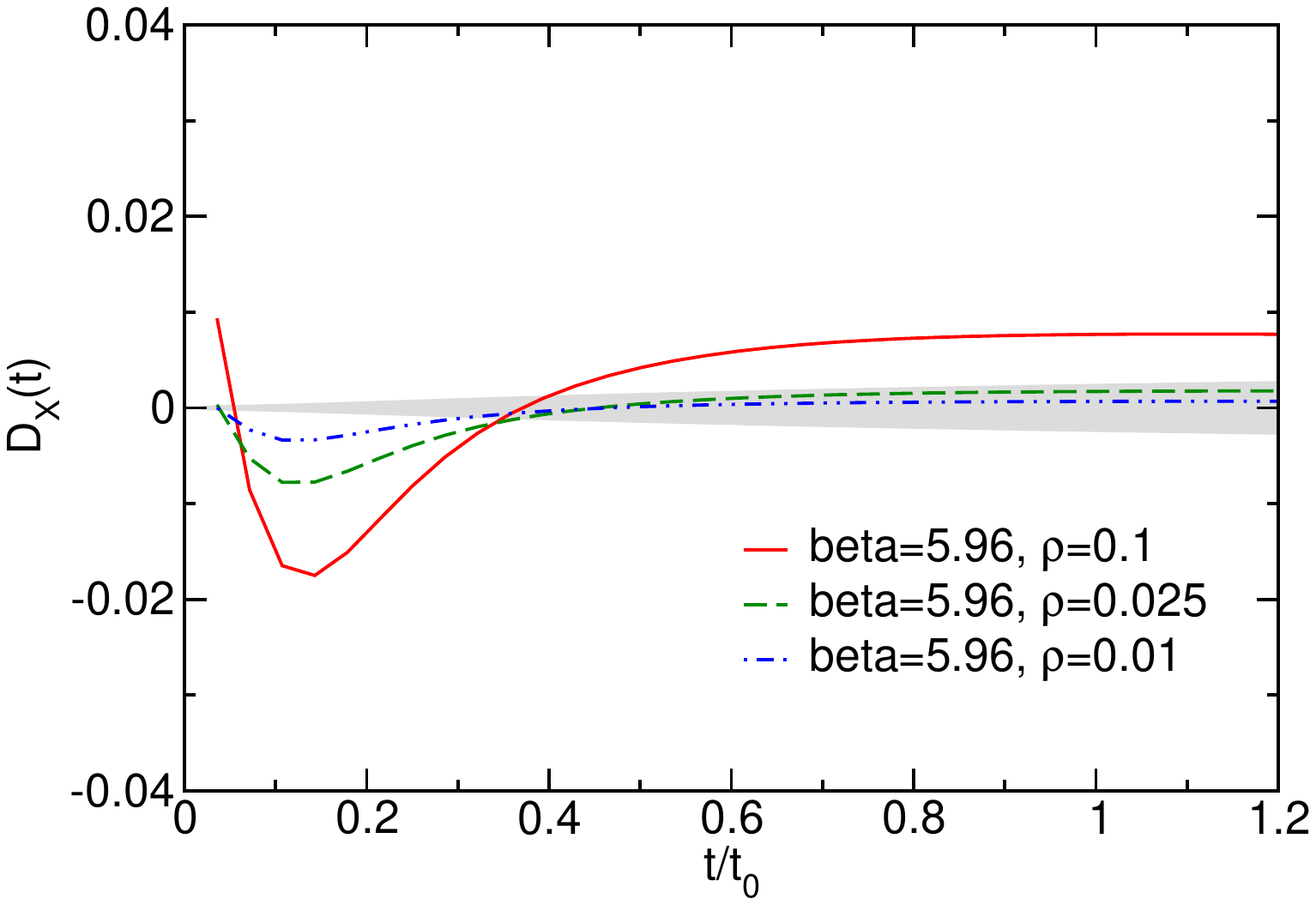} 
\caption{The behavior of 
$D_X(t)$ obtained with three smearing parameters:
$\rho=0.1$ (solid curve), $\rho=0.025$ (dashed curve),
and $\rho=0.01$ (double-dotted curve) 
as functions of $t/t_0$ calculated at $\beta=5.96$.
The area shaded in gray corresponds to the relative size of the statistical uncertainties on $X(t)$ obtained by the Wilson flow. }
\label{fig:X_b596}
\end{figure*}

In Fig.~\ref{fig:X_b596}, we next show the smearing parameter dependence of $D_X(t)$
calculated at $\beta=5.96$ as a typical example. The behavior of $D_X(t)$ is 
plotted as a function of $t/t_0$ for three smearing parameters: $\rho=0.01$ (solid curve),
$\rho=0.025$ (dashed curve), and $\rho=0.01$ (double-dotted curve). The relative 
difference of $X(t)$ between the Wilson flow and the stout-link smearing is smaller
than 2\% even for the case of $\rho=0.1$. When the smearing parameter decreases, the relative difference steadily get smaller as expected. For the stout-link smearing 
results obtained with the smaller smearing parameter ($\rho \le 0.025$), their deviations from the Wilson flow result becomes smaller than the statistical uncertainties on 
$X(t)$ obtained from the Wilson flow in the large-$t$ region ($t/t_0 > a^2/t_0\approx 0.358$).
We will also show the difference between two methods for
determining another scale $\omega_0$ in the Appendix.

\begin{figure*}[ht]
\centering
\includegraphics*[width=.8\textwidth]{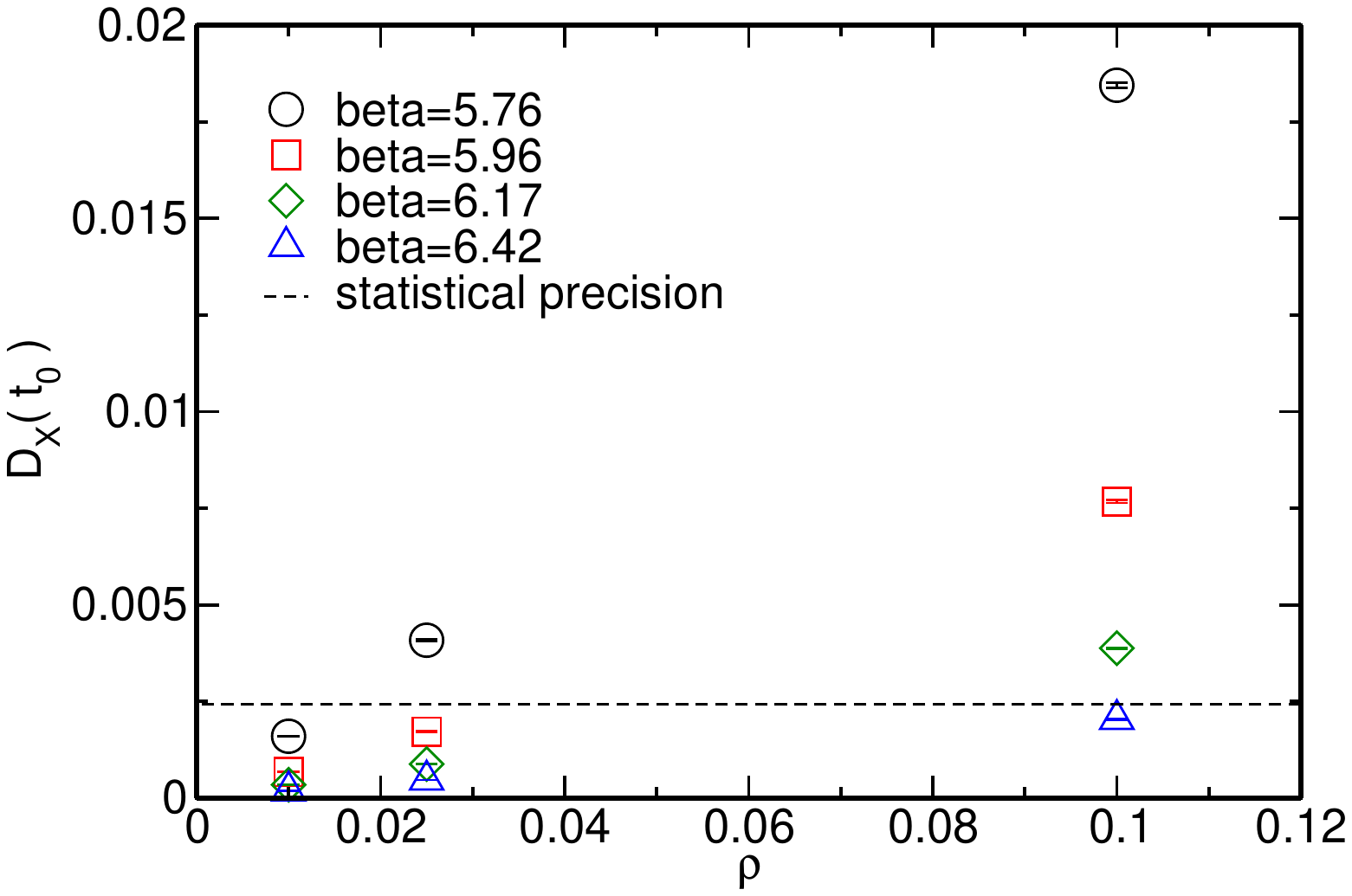} 
\caption{ 
The smearing parameter dependence of $D_X(t)$ evaluated at $t=t_{0}$
with three smearing parameters: $\beta=5.76$ (circles) ,
$\beta=5.96$ (squares), $\beta=6.17$ (diamonds), $\beta=6.42$ (upper triangles). 
A horizontal dotted line represents the relative size of the statistical uncertainties ($\sim 0.24\%$) on
the values of $t_0$ calculated by the Wilson flow in this study.}
\label{fig:diff_smear}
\end{figure*}

From above observations, we can expect that the Wilson flow and the stout-link smearing remain equivalent even with the finite parameters ($a\neq 0$ and $\rho\neq 0$) to determine the Wilson flow scale $t_0$ and $\omega_0$ within certain numerical precision. 
All values of $t_0$ and $\omega_0$ determined in this work are tabulated in 
Table~\ref{table:rho_dep_on_t0} for the scale $t_0$ and Table~\ref{table:rho_dep_on_w0} for the scale $\omega_0$. 
As for the scale $t_0$, we also plot the values of $D_X(t_0)$, which correspond to the relative differences between the Wilson flow and the stout smearing results, as a function of $\rho$ at four lattice spacings ($\beta=5.76$,
5.96, 6.17 and 6.42) in Fig.~\ref{fig:diff_smear}.
A horizontal dotted line represents the relative size of the statistical uncertainties on the values ($\sim 0.24\%$) of $t_0$ determined by the Wilson flow 
in this study. 

Figure~\ref{fig:diff_smear}
shows that the difference between the two methods at the coarse lattice spacing ($\beta=5.76$) becomes smaller than the statistical error when the smearing parameter is set as $\rho=0.01$. 
On the other hand, for the fine lattice spacing ($\beta=6.42$), the larger value of $\rho=0.1$ is enough to reproduce the Wilson flow result within the statistical precision. 
It is also worth emphasizing that the values of $D_X(t_0)$ seem to linearly depend on $\rho$ with fixed lattice spacing and vanish in the limit of $\rho\to0$ with
finite lattice spacings.
These observations are expected since the continuous version of the stout-smearing procedure is equivalent to the Wilson flow equation up to ${\cal O}(\rho)$ even at finite lattice spacing as shown in Sec.~\ref{sec:FLOW_STOUT}. 

\begin{figure*}[ht]
\centering
\includegraphics*[width=.8\textwidth]{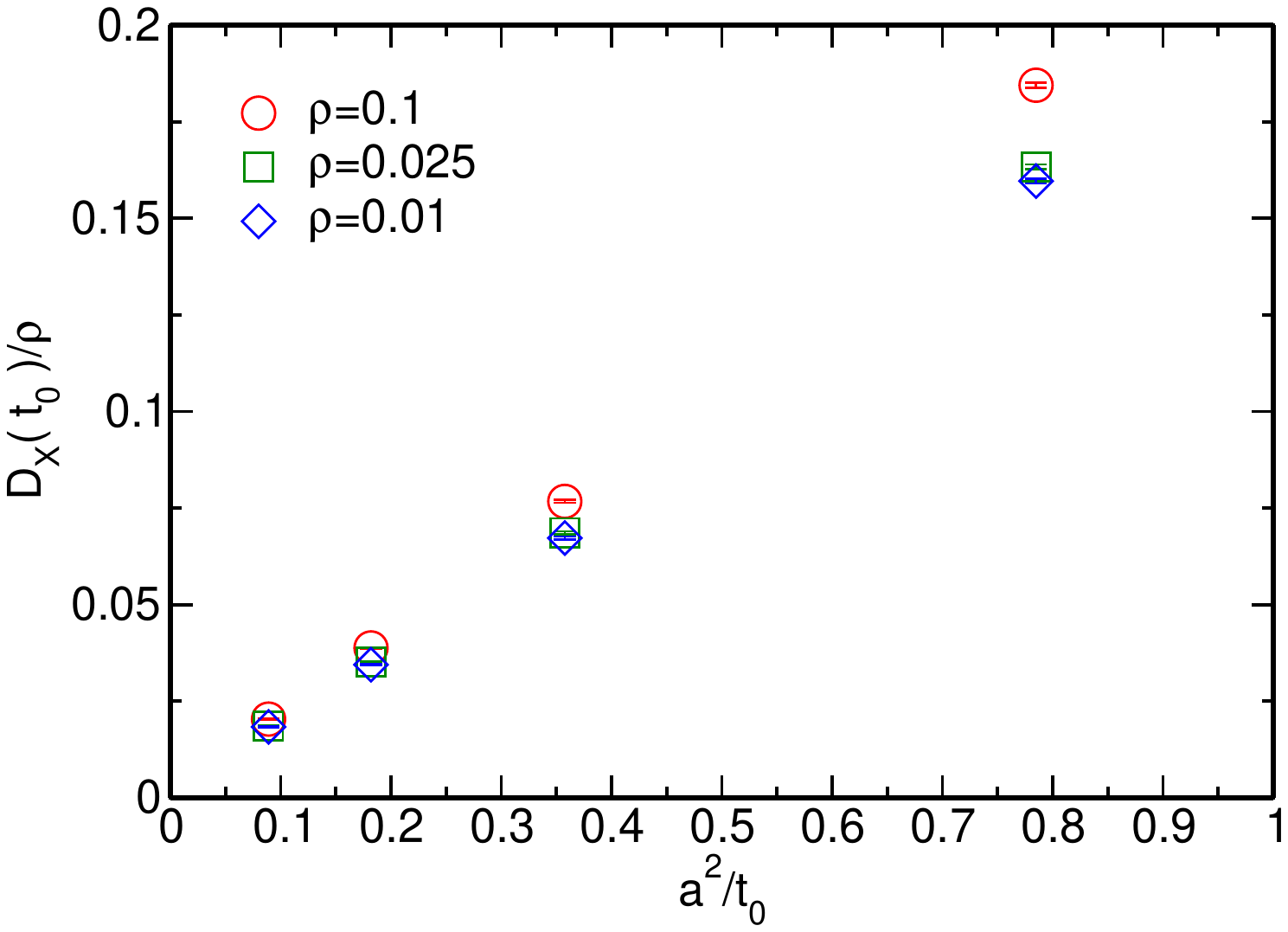} 
\caption{ 
The lattice spacing dependence of $D_X(t)$ divided by $\rho$ evaluated at $t=t_{0}$
with three smearing parameters: $\rho=0.1$ (circles) ,
$\rho=0.025$ (squares), and $\rho=0.01$ (diamonds). 
}
\label{fig:diff_spacing}
\end{figure*}

Moreover, the slope of the linear $\rho$ dependence has some lattice spacing dependence, which shows
that the slope becomes steeper as the lattice spacing increases. 
Indeed, the values of $D_X(t_0)$ seem to linearly depend on $a^2$ with fixed smearing parameter $0<\rho<1$ as shown in Fig~\ref{fig:diff_spacing}, where we plot the values of $D_X(t_0)/\rho$ as a function of $a^2/t_0$ with three smearing parameters ($\rho=0.1$, 0.025, and 0.01). Since the higher order corrections with respect to
$\rho$ become non-negligible for the larger value of $\rho$, the scaling of $D_X(t_0)/\rho$ as 
a function of $a^2/t_0$ get worse at the coarser lattice spacing. 
What we observe here is consistent with our expectation discussed in Sec~\ref{sec:FLOW_STOUT}, that
the leading order corrections on the equivalence of the two methods start at ${\cal O}(a^2)$ when the smearing parameter $\rho$ remains finite.

Next, it is worth remarking that the computational cost for the Wilson flow is 
relatively higher than the stout smearing. In our actual numerical code,
we find that the stout-link smearing is roughly a factor of ${\cal O}(10)$ 
faster than the Wilson flow even with the same numbers of flow iterations $n_{\rm flow}$
and smearing steps $n_{\rm st}$ as summarized in Table~\ref{tab:cpu_time}.
Although the required number of flow iterations increases quadratically 
as the lattice spacing decreases, the Wilson flow can be replaced by 
the stout smearing at the finer lattice spacing and then the computational cost
can be significantly reduced. This is just an application in terms of the 
effectiveness of numerical calculations.

On the other hand, there is a reverse application of this equivalence.
The stout smearing procedure can be implemented in the Hybrid Monte Carlo based updating algorithms because of its differentiability with respect to the link variables. Recently, the stout smearing is partly used in the definition of the lattice fermion action. Therefore, some technique for calculating one loop quantities in lattice perturbation theory is required to be developed for the smeared-link fermion actions. 
The findings of this study warrants the use of perturbation calculations 
in the gradient flow formalism for such purpose. 

%
%
%

%
%
%

%
%

%
%

%
%
\begin{table*}[ht]
\caption{
Results of $t_{0}/a^2$ obtained from the stout smearing with three smearing parameters $\rho=0.1,~0.025,~0.01$ and the Wilson flow ($\epsilon=0.025$). 
\label{table:rho_dep_on_t0}
}
\begin{ruledtabular} 
\begin{tabular}{lccccccc}
\hline
 &\multicolumn{3}{c}{$t_{0}/a^2$ (stout-link smearing)} & $t_{0}/a^2$ (Wilson flow) \cr
 \cline{2-4}
 $\beta$ & $\rho=0.1$ & $\rho=0.025$ & $\rho=0.01$ & $\epsilon=0.025$\cr
\hline
5.76 & 1.2502(30) & 1.2690(30) & 1.2722(31) & 1.2741(31)\cr
5.96 & 2.7744(62) & 2.7919(62) & 2.7949(62) & 2.7968(62)\cr
6.17 & 5.476(13)  & 5.494(13) & 5.497(13)   & 5.499(13)\cr
6.42 & 11.218(22) & 11.236(23) & 11.240(23) & 11.242(23)\cr
\hline
\end{tabular}
\end{ruledtabular}
\end{table*}
%

%
%
\begin{table*}[ht]
\caption{
Results of $w_{0}/a$ obtained from the stout-link smearing with three smearing parameters $\rho=0.1,~0.025,~0.01$
and the Wilson flow ($\epsilon=0.025$).
\label{table:rho_dep_on_w0}
}
\begin{ruledtabular} 
\begin{tabular}{lccccccc}
\hline
 &\multicolumn{3}{c}{$w_{0}/a$ (stout-link smearing)} & $w_{0}/a$ (Wilson flow) \cr
 \cline{2-4}
 $\beta$ & $\rho=0.1$ & $\rho=0.025$ & $\rho=0.01$ & $\epsilon=0.025$\cr
\hline
5.76 & 1.1098(18) & 1.1199(18) & 1.1220(18) & 1.1224(18)\cr
5.96 & 1.6755(24) & 1.6819(24) & 1.6832(24) & 1.6833(24)\cr
6.17 & 2.3684(41) & 2.3729(41) & 2.3738(41) & 2.3738(41)\cr
6.42 & 3.4042(48) & 3.4075(48) & 3.4081(48) & 3.4081(48)\cr
\hline
\end{tabular}
\end{ruledtabular}
\end{table*}
%

%
%
\begin{table*}[ht]
\caption{
Comparison of CPU execution time between the stout-link smearing
and the Wilson flow applying for full space-time 
dimensions and only spatial dimensions
with fixed $n_{\rm st}=n_{\rm flow}=100$ steps using a single configuration for the lattice size of $16^3\times 32$
(run on a single core of Intel Xeon E5-2609 CPU).
For the Wilson flow, we examine the second-order, third-order, and 
forth-order Runge-Kutta (RK) schemes, respectively. 
\label{tab:cpu_time}
}
\begin{ruledtabular} 
\begin{tabular}{ccccc}
\hline
directions &stout-link smearing ($\rho=0.025$)
 &\multicolumn{3}{c}{Wilson flow ($\epsilon=0.025$)} \cr
 \cline{3-5}
 & & 2nd-order RK & 3rd-order RK & 4th-order RK \cr
\hline
space-time & 136.80 [sec] & 985.55 [sec]  & 1496.51 [sec] & 2061.08 [sec] \cr 
space & 90.39 [sec] & 511.84 [sec]  & 763.51 [sec] & 1074.07 [sec] \cr 
\hline
\end{tabular}
\end{ruledtabular}
\end{table*}

\clearpage
\section{Summary} 
\label{sec:SUMMARY}

We have studied the equivalence between the Wilson flow and the stout smearing, which are analytically proven at finite lattice spacing $a$ in the zero limit of the stout-smearing
parameter $\rho$.
To demonstrate the equivalence of both methods by directly comparing the expectation value of the action density, numerical simulations have been performed with
the Wilson gauge configurations generated at four different gauge couplings ($\beta=5.76$, 5.96, 6.17, and 6.42). It is found that the two methods remain equivalent within some numerical precision even with finite parameters ($a\neq 0$ and $\rho\neq 0$) if the proper combination of two parameters ($a$ and $\rho$) is chosen. Especially, we verified that $\rho=0.1$ for $\beta=6.42$, $\rho=0.025$ for $\beta=5.96$ and 6.17, and $\rho=0.01$ for $\beta=5.76$ are enough small to identify the stout smearing with the Wilson flow in the determination of the reference scales, $t_0$ and $\omega_0$. 

Note that the computational cost of the stout-link smearing is roughly a factor of ${\cal O}(10)$ lower than the Wilson flow even with the same numbers of flow iterations $n_{\rm flow}$ and smearing steps $n_{\rm st}$. Therefore, the Wilson flow can be potentially replaced by the stout smearing at the finer lattice spacing and then the computational cost can be significantly reduced. Vise verse, we consider that the findings of this study warrants the use of perturbation calculations developed in the gradient flow formalism in order to calculate one loop quantities in lattice perturbation theory for the smeared-link fermion action. 

\begin{acknowledgments}

We would like to thank M. Ammer and R.J. Hudspith for useful discussions.
K. S. is supported by Graduate Program on Physics for the Universe (GP-PU)
of Tohoku University. 
Numerical calculations in this work were partially performed using Yukawa-21 
at the Yukawa Institute Computer Facility,
and also using Cygnus at Center for Computational Sciences (CCS), University of Tsukuba
under Multidisciplinary Cooperative Research Program of CCS (MCRP-2021-54, MCRP-2022-42). 
This work was also supported in part by Grants-in-Aid for Scientific Research form the Ministry 
of Education, Culture, Sports, Science and Technology (No. 18K03605 and No. 22K03612).
\end{acknowledgments}
\appendix
\section*{Appendix: Difference in determination of the scale $\omega_0$}
\label{appendix}

Let us introduce $W(t)$, which is defined as $W(t)=t\frac{d}{dt}\left<t^2E(t)\right>$. 
In order to measure the difference between two methods for determining another scale $\omega_{0}$, we then calculate the following ratio:
%
%
\begin{equation}
D_W(t)=\frac{W_{\rm{stout}}(t)-W_{\rm{flow}}(t)}{W_{\rm{flow}}(t)}
\end{equation}
where $W_{\rm stout}(t)$ denotes the corresponding quantity of $W(t)$ calculated
by using the stout-link smearing as a function of $t=\rho n_{\rm st}$, while
$W_{\rm flow}(t)$ is calculated by the Wilson flow. 

\begin{figure*}[ht]
\centering
\includegraphics*[width=.8\textwidth]{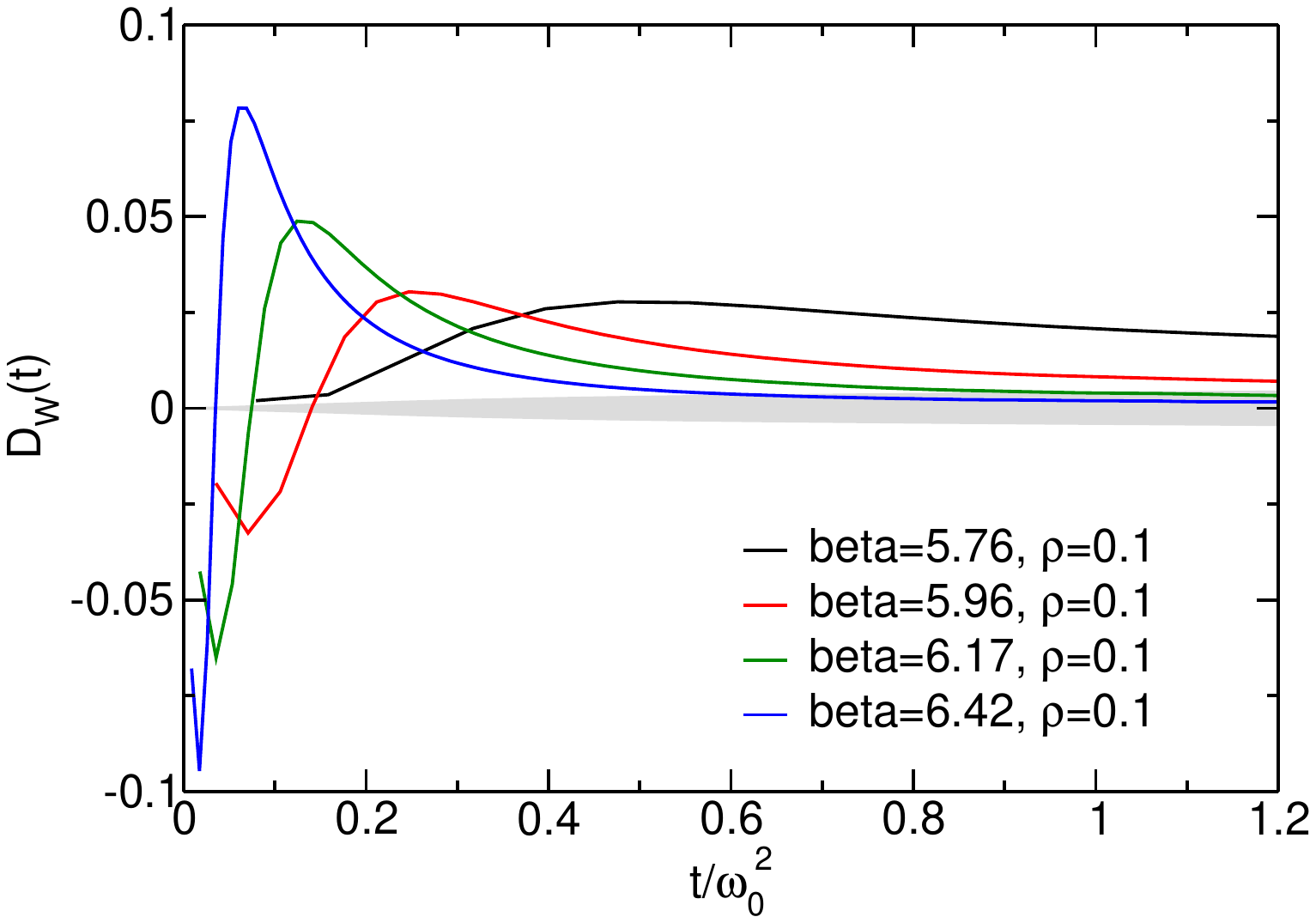} 
\caption{
The lattice spacing dependence of $D_W(t)$ calculated between the Wilson flow and the stout smearing ($\rho=0.1$) at $\beta=5.76$, 5.96, 6.17, and 6.42.
The area shaded in gray corresponds to the relative size of the statistical uncertainties
on $W(t)$ obtained by the Wilson flow. }
\label{fig:W_r01}
\end{figure*}

\begin{figure*}[ht]
\centering
\includegraphics*[width=.8\textwidth]{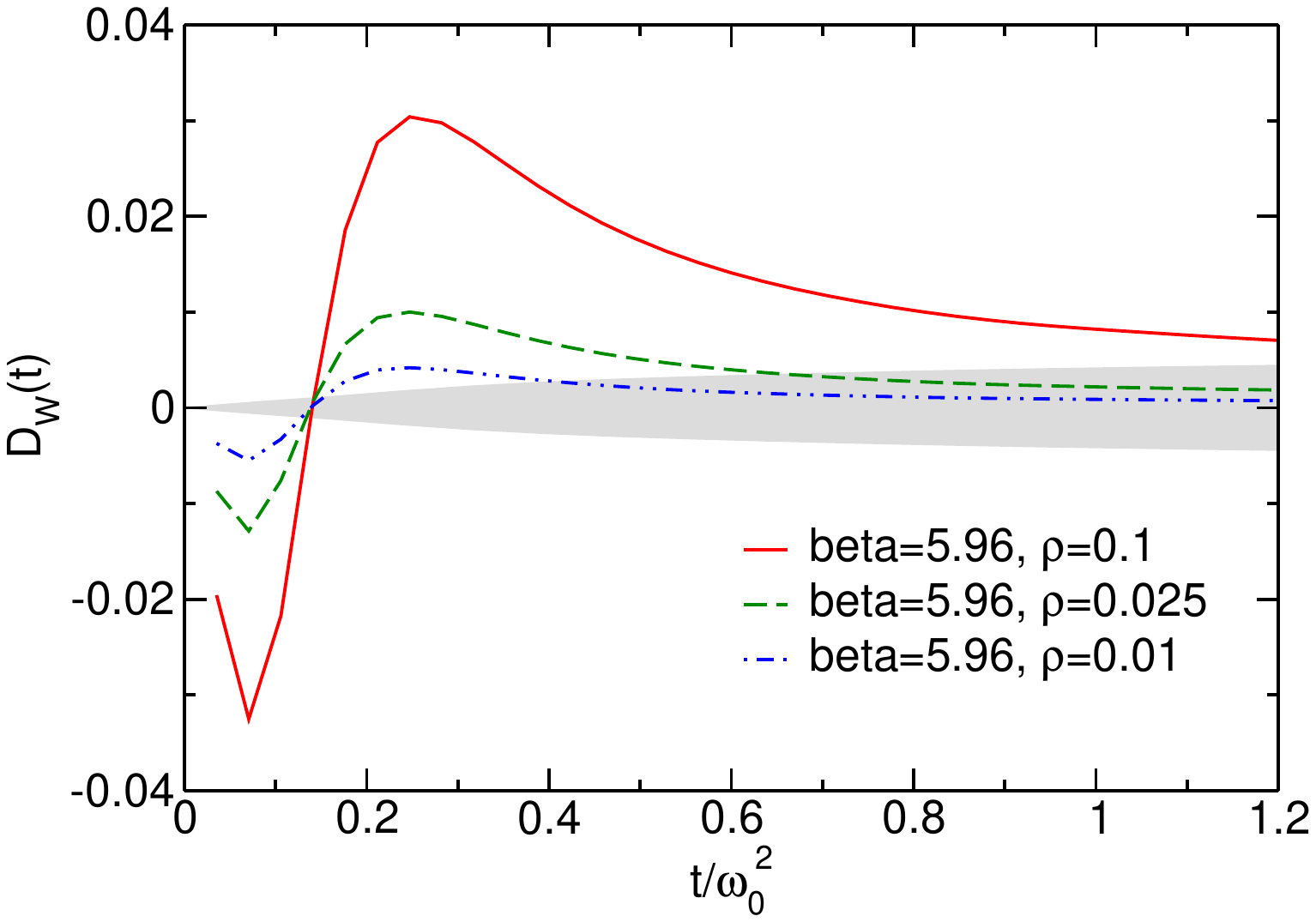} 
\caption{
The behavior of $D_W(t)$ obtained with three smearing parameters:
$\rho=0.1$ (solid curve), $\rho=0.025$ (dashed curve),
and $\rho=0.01$ (double-dotted curve) 
as functions of $t/t_0$ calculated at $\beta=5.96$.
The area shaded in gray corresponds to the relative size of the statistical uncertainties on $W(t)$ obtained from the Wilson flow. }
\label{fig:W_b596}
\end{figure*}

Figure~\ref{fig:W_r01} shows the lattice spacing dependence of $D_W(t)$ calculated between the Wilson flow and the stout smearing ($\rho=0.1$) at $\beta=5.76$, 5.96, 6.17, and 6.42, while Figure~\ref{fig:W_b596} shows the smearing parameter dependence of $D_W(t)$ 
calculated at $\beta= 5.96$. In similar to the case of $D_X(t)$, 
the relative differences are barely saturated at $t \approx \omega_0^2$, which is located in
the large-$t$ region ($t/\omega_0^2 > a^2/\omega_0^2$). 

\end{document}